\documentclass[runningheads]{llncs}
\usepackage{amsfonts}

\usepackage{latexsym}
\usepackage{amssymb}
\usepackage{epsfig}
\usepackage{amsmath}
\usepackage[mathscr]{eucal}
\usepackage{subfigure}
\usepackage{graphicx}

\usepackage[linesnumbered,ruled,vlined]{algorithm2e}

\SetKwInOut{Input}{Input}
\SetKwInOut{Output}{Output\,}
\SetKwInOut{Data}{Data}
\SetKwProg{SequentialDecode}{SequentialDecode}{}{EndSequentialDecode}
\SetKwProg{SeqDecodeMod}{SeqDecodeMod}{}{EndSeqDecodeMod}
\SetKwProg{CorrectDeletion}{CorrectDeletion}{}{EndCorrectDeletion}
\SetKwProg{CorrectFlip}{CorrectFlip}{}{EndCorrectFlip}
\SetKwProg{CorrectErasure}{CorrectErasure}{}{EndCorrectErasure}
\SetKwProg{BurstDecode}{BurstDecode}{}{EndBurstDecode}
\SetKwProg{OrdDecode}{OrdDecode}{}{EndOrdDecode}
\SetKwProg{MultDecode}{MultDecode}{}{EndMultDecode}
\SetKwProg{MultDelErrDecode}{MultDelErrDecode}{}{EndMultDelErrDecode}

\makeatletter
\def\BState{\State\hskip-\ALG@thistlm}
\makeatother

\newcommand{\ind}{1\hspace{-2.3mm}{1}}

\begin{document}

\title{Real-time error correction codes for deletable errors}
\titlerunning{Error correction for deletable errors}
\author{ \textbf{Ghurumuruhan Ganesan}}
\authorrunning{G. Ganesan}
\institute{Institute of Mathematical Sciences, HBNI, Chennai\\
\email{gganesan82@gmail.com }}

\date{}
\maketitle

\begin{abstract}
In this paper we study codes for correcting deletable errors in binary words, where each bit is either retained, substituted, erased or deleted and the total number of errors is much smaller compared to the length of the codeword. We construct codes capable of correcting errors in the received codeword in real-time with small delay and low probability of decoding error.

\vspace{0.1in} \noindent \textbf{Key words:} Deletable errors, redundancy of codes, real-time correction.

%\vspace{0.1in} \noindent \textbf{AMS 2000 Subject Classification:} Primary: 94B60, 94B70.
\end{abstract}

\bigskip

\setcounter{equation}{0}
\renewcommand\theequation{\arabic{section}.\arabic{equation}}
\section{Introduction} \label{intro}
Consider recovery of a binary word corrupted by deletable errors, where each error could either be an erasure, bit substitution (i.e., a one is converted to a zero and vice versa) or deletion. Regarding deletions,~\cite{lev} used Varshamov-Tenengolts codes (VT codes) for correcting a single deletion. Later~\cite{helb} described codes for correcting arbitrary number of deletions and recently, in~\cite{sima}, codes with near optimal redundancy for combating fixed number of deletions are constructed using hash functions. Error correction codes with time constraints have also been studied before. For example~\cite{rmj} focused on developing bidirectional interactive algorithms with low information exchange for correcting deletions. An estimate on the number of rounds needed for synchronization was obtained using a divide and conquer strategy. Streaming codes with delay constraints for erasure and burst channels have been studied in~\cite{krishnan} (see also references therein).

%Indeed Levenshtein (1965) showed how to correct a single deletion using (VT) codes introduced in can be used to correct a single deletion, no matter the location of the deleted bit within the word. %Since then many improvements and algorithms have been proposed for various channel modes including burst deletions (Schoeny et al 2017), communication complexity (Orlitsky (1993)) and multiple insertions/deletions (Helberg et al (2002)).

%WRITE MORE ON THE DELETABLE ERRORS...

%CHGN BERLOWE +ETC...

In this paper, we seek codes with low redundancy that are capable of correcting deletable errors (that could be an erasure, substitution or deletion)
in real-time with low delay, a notion made precise in the next section. We use modified Varshamov-Tenengolt codes to construct the desired codes and also obtain an estimate on the fraction of error patterns that are correctable by the proposed codes.

The paper is organized as follows: In Section~\ref{state} we state our main result Theorem~\ref{thm_12} regarding codes capable of real-time correction with small delay and also Theorem~\ref{thm_11} describing codes for correcting~\(P-\)far deletable error patterns. In Section~\ref{p_far_codes}, we prove Theorem~\ref{thm_11} and also describe an error correction algorithm~\({\cal A}\) for correcting~\(P-\)far errors. In Section~\ref{pf_12}, we use the correction algorithm~\({\cal A}\) to construct codes with real-time correction capability and prove Theorem~\ref{thm_12}.

%In Section~\ref{prelim}, we study a special class of error patterns called~\(P-\)far error patterns used in the proof of Theorem~\ref{thm_12}. In~Section~\ref{pf2}, we prove Theorem~\ref{thm_12} and  redundancy upper bounds for codes capable of correcting~\(P-\)far deletable error patterns. In Section~\ref{pf_red_min}, we prove the redundancy lower bounds for codes capable of correcting~\(P-\)far error patterns and in Section~\ref{appendix}, we have the Appendix collecting auxiliary results used in the proofs.

%This can be done for example by having a markers placed within the transmitted codeword

\setcounter{equation}{0}
\renewcommand\theequation{\arabic{section}.\arabic{equation}}
\section{Main results}\label{state}
A word of length~\(n\) is a vector~\(\mathbf{x} = (x_1,\ldots,x_n) \in \{0,1\}^n.\) A deletable error pattern of length~\(n\) is an element~\(\mathbf{g} = (\mathbf{e},\mathbf{v})  = ((e_1,\ldots,e_n),(s_1,\ldots,s_n))\in \{0,1\}^{n} \times \{D,E,F\}^{n},\) with~\(D, E\) and~\(F\) denoting deletion, erasure and substitution, respectively. The word~\(\mathbf{y} = F_{\mathbf{g}}(\mathbf{x})\) obtained after~\(\mathbf{x}\) is corrupted by the error pattern~\(\mathbf{g}\) is defined as follows: If~\(e_i = 1,\) then the bit~\(x_i\) is deleted, erased or substituted depending on whether~\(s_i = D, E\) or~\(F,\) respectively. If~\(e_i = 0,\) no change occurs to the bit~\(x_i.\) For integer~\(1 \leq t \leq n\) we define~\(F_{\mathbf{e}}(\mathbf{x},t)\) to be the result of corrupting the first~\(t\) bits of~\(\mathbf{x}\) according to the first~\(t\) entries in~\(\mathbf{e}.\)

For example, if~\(n = 5\) and~\(\mathbf{g} = ((1,0,1,1,1),(F,E,D,F,E)),\) then~\(F_{\mathbf{g}}(\mathbf{x}) = (1-x_1,x_2,1-x_4,\varepsilon),\) where~\(\varepsilon\) is the erasure symbol and~\(F_{\mathbf{g}}(\mathbf{x},3) = (1-x_1,x_2).\)
For~\(r \geq 1,\) we define~\({\cal E}_n(r) := \left\{(\mathbf{e},\mathbf{v}) : \sum_{i=1}^{n} e_i \leq r \right\}\) to be the set of all possible~\(n-\)length patterns containing at most~\(r\) deletable errors. We also let~\({\cal E}_n := \cup_{1 \leq r \leq n} {\cal E}_n(r)\) be the set of all possible~\(n-\)length error patterns.

An~\(n-\)length code~\({\cal C}\) of size~\(\#{\cal C} = q\) is a set~\(\{\mathbf{x}_1,\ldots,\mathbf{x}_q\} \subset \{0,1\}^{n}.\)
We define the redundancy of~\({\cal C}\) to be
\begin{equation}\label{red_def}
R({\cal C}) := n - \log\left(\#{\cal C}\right),
\end{equation}
where all logarithms are to the base~\(2\) throughout.

%there exists a map
%\[g = g_{\cal F} : \bigcup_{\mathbf{x} \in {\cal C}} \bigcup_{\mathbf{e} \in {\cal F}} F_{\mathbf{e}}(\mathbf{x}) \longrightarrow {\cal C}.\] In other words, %Let~\({\cal C}\) be any code and for~\(1 \leq i \leq n,\) let~\({\cal C}(i)\) be the code formed by the set of all codewords obtained by taking the first~\(i\) bits in each codeword of~\({\cal C}.\) For~\(\mathbf{x} \in {\cal C}\) and an error pattern~\(\mathbf{e}\) we recall that~\(\mathbf{y} := F_{\mathbf{e}}(\mathbf{x})\) is the word obtained by corrupting~\(\mathbf{x}\) according to the error pattern~\(\mathbf{e}.\)

Let~\({\cal F} \subseteq {\cal E}_n\) be any set of~\(n-\)length error patterns. The code~\({\cal C}\) is said to be capable of correcting all error patterns in~\({\cal F}\)  if for any~\(\mathbf{x}_1 \neq \mathbf{x}_2 \in {\cal C}\) and any~\(\mathbf{g}_1,\mathbf{g}_2 \in {\cal F},\) we must have that the corruptions~\(F_{\mathbf{g}_1}(\mathbf{x}_1) \neq F_{\mathbf{g}_2}(\mathbf{x}_2).\) We have the following definition regarding real-time error correction.
\begin{definition}\label{real_time}
For integer~\(d \geq 1\) we say that~\({\cal C}\) can correct all error patterns in~\({\cal F} \subseteq {\cal E}_n\) in real-time with delay at most~\(d\) if for any~\(1 \leq z \leq n-d, \mathbf{g}_1,\mathbf{g}_2 \in {\cal F}\) and ~\(\mathbf{x}_1\neq \mathbf{x}_2 \in {\cal C},\) we have that
\[F_{\mathbf{g}_1}(\mathbf{x}_1,z+d) \neq F_{\mathbf{g}_2}(\mathbf{x}_2,z+d)\]
\end{definition}
The consequence of Definition~\ref{real_time} is this: Suppose the word~\(\mathbf{x}\) is transmitted and is corrupted by the error pattern~\(\mathbf{g}.\) Given the partial word~\(F_{\mathbf{g}}(\mathbf{x},z+d)\) obtained by corrupting the first~\(z+d\) bits of~\(\mathbf{x}\) for any~\(1 \leq z \leq n-d,\) we must be able to determine the first~\(z\) bits of~\(\mathbf{x}.\)

For integers~\(t,d \geq 1\) let~\({\cal F}(t,d) \subset {\cal E}_n(t)\) be a set of maximum size containing at most~\(t\) errors, that are correctable by the code~\({\cal C}\) in real-time with delay~\(d.\) We define the~\((t,d)-\)\emph{real-time error probability} of~\({\cal C}\) to be
\[\rho(t,d) := \frac{\#\left({\cal E}_n(t) \setminus {\cal F}(t,d)\right)}{\#{\cal E}_n(t)}.\] If all error patterns in~\({\cal E}_n(t)\) are equally likely, then~\(\rho(t,d)\) could be interpreted as the decoding error probability.

For a given~\(t,\) we are interested in studying the tradeoff involved in choosing the delay parameter~\(d.\) If~\(d=\infty,\) then we simply require a code that corrects~\(t\) errors and we know (see for e.g.~\cite{sima}) there are codes with negligible redundancy for correcting any fraction of errors.  Intuitively therefore if~\(d\) is large, then our code should have small redundancy since the effect of the delay becomes lesser and lesser as~\(d\) increases. What about the redundancy when~\(d\) is small? In the extreme case of~\(d=1,\) we must simply have a code that is equivalent to a repetition code since the first bit must be deductable using the second bit and the first two bits must be inferred using the first three bits and so on. In this case, the redundancy is~\(1-\frac{1}{n}.\) In general, for small~\(d\) therefore we expect the code to have a large redundancy.

%For~\(n \geq 1\) let~\({\cal C}\) be a~\(n-\)length code and~\(1 \leq t_n \leq n\) be any integer. Recall from discussion prior to~(\ref{en_def}) that~\({\cal E}_n(t_n)\) is the set of all possible error patterns containing at most~\(t_n\) deletable errors. We say that~\({\cal C}\) corrects at least a fraction~\(f \in (0,1)\) of patterns in~\({\cal E}_n(t_n)\) if there exists~\({\cal F} \subseteq {\cal E}_n(t_n)\) such that~\({\cal C}\) is~\({\cal F}-\)correcting and~\(\#{\cal F} \geq f \cdot \#{\cal E}_n(t_n).\)

The following result obtains bounds on the delay, redundancy and the error probability of codes capable of correcting deletable errors, as a function of the total number of errors.
\begin{theorem}\label{thm_12} Let~\(1 \leq t_n \leq \frac{n^{\frac{1}{3}}}{6}\) be any integer. For every integer~\(d_n\) satisfying~\(4t_n \leq d_n \leq \frac{2n}{3t_n^2}\) there is an~\(n-\)length code~\({\cal C}_{frac} = {\cal C}_{frac}(n)\) with a real-time delay~\(d_n,\)
redundancy
\begin{equation}\label{red_upp}
R({\cal C}_{frac}) \leq n \cdot \frac{4\log(2d_n)}{d_n}, %chck CHNG HERE...
\end{equation}
and a~\((t_n,d_n)-\)real-time error probability
\begin{equation}\label{err_prob}
\rho(t_n,d_n) \leq 11\frac{t_n^2d_n}{n}.
\end{equation}
\end{theorem}
Thus if the maximum number of deletable errors in an~\(n-\)length word is~\(t_n,\) then for every~\(d_n\) satisfying the bounds in the Theorem, we can obtain a code capable of real-time correction with delay at most~\(d_n,\) redundancy as in~(\ref{red_upp}) and total error probability as in~(\ref{err_prob}). Ideally we prefer that~\(d_n \longrightarrow \infty\) and~\(\frac{t_n^2 d_n}{n} \longrightarrow 0\) as~\(n \rightarrow \infty\) so that for all~\(n\) large the code~\({\cal C}_{frac}\) has redundancy~\(o(n)\) and an error probability of~\(o(1),\) using the standard~\(o(.)\) notations. Theorem~\ref{thm_12} also illustrates the redundancy-error probability tradeoff in terms of the desired delay~\(d_n.\) Choosing larger~\(d_n\) reduces redundancy but increases the error probability. We illustrate our result with some examples.\\\\
\underline{\emph{Example 1}}: Suppose we have a communication channel that introduces at most~\(n^{\alpha}\) deletable errors in an~\(n-\)length word for some constant~\(\alpha < \frac{1}{3}.\) Setting~\(t_n = n^{\alpha},\) choosing~\(d_n = 4n^{\alpha}\cdot \log{n}\) and using~\(\alpha < \frac{1}{3},\) we see that the conditions in Theorem~\ref{thm_12} are satisfied for all~\(n\) large. From Theorem~\ref{thm_12} we therefore see that there exists an~\(n-\)length code with real-time delay~\(d_n,\) redundancy~\(o(n^{1-\alpha})\) and a real-time error probability of at most~\(\frac{42 \log{n}}{n^{1-3\alpha}}.\)\\\\
\underline{\emph{Example 2}}: For a numerical example, suppose that an encoded file containing~\(n = 10^{8}\) bits is to be transmitted across a channel that introduces at most~\(t_n = 10\) deletable errors.  Setting~\(d_n = 4000\) we see that the redundancy of the code~\({\cal C}_{frac}\) is at most~\(10^{6} \cdot \log(2000) \leq 12 \cdot 10^{6}\) and so the rate of~\({\cal C}_{frac}\) is at least~\(1-\frac{12 \cdot 10^{6}}{10^{8}} = 0.88.\) If all error patterns are equally likely then the real-time error probability is also the decoding error probability and is at most~\(\frac{44}{10000} = 4.4 \cdot 10^{-3}.\)

The main ingredient in the construction of the real-time error correction code in Theorem~\ref{thm_12} is~\(P-\)far deletable error patterns which we describe next. Let~\(n \geq P \geq 2\) be any integers. If~\(\mathbf{g} =(\mathbf{e},\mathbf{v}), \mathbf{e} = (e_1,\ldots,e_n) \in \{0,1\}^{n}\) is an error pattern such that~\(|i-j| \geq P\) for any two non zero~\(e_i,e_j,\) we then say that~\(\mathbf{g}\) is an \(n-\)length \(P-\)\emph{far deletable error pattern}.
Letting
\begin{equation}\label{del_def_main}
\delta(P) := \frac{2P+1}{2^{P-1}},
\end{equation}
we have the following result regarding codes capable of correcting~\(3P-\)far deletable error patterns.
\begin{theorem}\label{thm_11} For all integers~\(n \geq P \geq 2\) there exists a \(n-\)length code~\({\cal C}_{far} = {\cal C}_{far}(n,P)\) with redundancy
\begin{equation}\label{red_upp44}
R({\cal C}_{far}) \leq \left(\frac{n}{P}-1\right)\log\left(\frac{2P+1}{1-\delta(P)}\right) + \log{P} + 2
\end{equation}
that is capable of correcting all~\(3P-\)far deletable error patterns.

Conversely, if~\({\cal A}\) is any \(n-\)length code capable
of correcting all~\(3P-\)far deletable error patterns, then
\begin{equation}\label{min_red2}
R({\cal A}) \geq  \frac{n}{64(3P+6)} - 3  %chk agn...for our bnft +ETc..
\end{equation}
for all large~\(n.\)

Suppose~\(\{P_n\}\) is a sequence satisfying
\begin{equation}\label{pn_cond}
\frac{P_n}{\log{n}} \longrightarrow \infty \text{ and } \frac{P_n}{\sqrt{n\log{n}}} \longrightarrow 0
\end{equation}
as~\(n \rightarrow \infty.\) If~\({\cal B}\) is any \(n-\)length code capable
of correcting all~\(3P_n-\)far deletable error patterns, then
\begin{equation}\label{min_red22}
R({\cal B}) \geq  \left(\frac{n}{6P_n}-1\right)\log\left(\frac{3P_n}{64}\right) - 2 %CHCK %CONSTs HERE...
\end{equation}
for all large~\(n.\)
\end{theorem}
For~\(P = P_n\) growing with~\(n\) we see from Theorem~\ref{thm_11}  that the redundancy grows at least of the order of~\(n \cdot \frac{\log{P_n}}{P_n} = o(n).\) For this range of~\(P,\) we also see from the upper bound~(\ref{red_upp44}) that the code~\({\cal C}_{far}\) has near optimal redundancy.

Before we proceed with the proofs, we briefly contrast and compare our results with related existing schemes for deletion correction.
In our proof of Theorem~\ref{thm_12} below, we use Theorem~\ref{thm_11} to first construct a block code that is capable of correcting all~\(P_n-\)far deletable errors and then choose~\(P_n\) appropriately so as to obtain the desired delay~\(d_n.\) The advantage of using~\(P_n-\)far deletable errors is that we can ensure that at most one deletable error occurs within any one block (defined formally in the next section). Such an approach is known as \emph{segmentation} and has been studied before for edit channels (i.e. channels allowing only deletion or insertion of bits). The Liu-Mitzenmacher construction~\cite{liu} employs a segment-by-segment code specified via sufficient conditions and finds a maximum size code by converting into an independent set problem. Recently,~\cite{abro} has studied coding for segmented edit channels by using subsets of Varshamov-Tenengolts (VT) codes and obtained bounds on the maximum possible size of such a code.

The approaches described above are not directly applicable for compound channels allowing deletable errors because, in the case of segmentation, we can no longer interpret a substitution as a deletion followed by insertion as this would cause two errors within a block. Below, we use a modified form of VT codes to obtain our desired codes and demonstrate how deletable error correction can be performed in a sequential manner.

In the next two sections, we prove Theorems~\ref{thm_11} and~\ref{thm_12} in that order.

\setcounter{equation}{0}
\renewcommand\theequation{\arabic{section}.\arabic{equation}}
\section{Proof of Theorem~\ref{thm_11}}\label{p_far_codes}
We begin with a description of codes capable of correcting~\(P-\)far errors.

\subsection*{Codes for correcting~\(P-\)far errors}
For integers~\(m \geq 3\) and~\(0 \leq a \leq m,\) we define the modified Varshamov-Tenengolts (VT) code~\(VT_{a}(m)\)
\begin{equation}\label{vt_def}
VT_{a}(m) := \left\{\mathbf{x} = (x_1,\ldots,x_m) \in \{0,1\}^{m} : \sum_{i=1}^{m} i \cdot x_i \equiv a \mod{2m+1}\right\}. \nonumber\\
\end{equation}
Arguing as in~\cite{tat}, the code~\(VT_a(m)\) is capable of correcting a single deletable error.

For integers~\(P \geq 2\) and~\(n \geq P,\) write~\(n = t P + s\) where~\(0 \leq s < P-1.\) Let~\(\mathbf{0}\) and~\(\mathbf{1}\) denote the all zero and all ones words with length depending on the context. For integers~\(0 \leq a_1 \leq 2P\) and~\(0 \leq a_2 \leq 2(P+s),\) define the code~\({\cal C}_{far} = {\cal C}_{far}(a_1,a_2,P,s,n)\) as
\begin{eqnarray}
{\cal C}_{far} &:=& \{\mathbf{x} = (\mathbf{x}(1),\ldots,\mathbf{x}(t)) : \mathbf{x}(i) \in VT_{a_1}(P) \setminus \{\mathbf{0},\mathbf{1}\}   \nonumber\\
&&\;\;\;\;\;\;\;\; \text{ for }1 \leq i \leq t-1, \mathbf{x}(t) \in VT_{a_2}(P+s)\},
\label{code_def2}
\end{eqnarray}
so that~\({\cal C}_{far}\) is obtained by appending together~\(t-1\) words from the code\\\(VT_{a_1}(P) \setminus \{\mathbf{0},\mathbf{1}\}\) and then appending a word containing~\(P+s\) bits from the code~\(VT_{a_2}(P+s).\) We denote the bits in the~\(i^{th}\) word as~\(\mathbf{x}(i) = (x_1(i),\ldots,x_P(i))\) for~\(1 \leq i \leq t-1\) and let~\(\mathbf{x}(t) = (x_1(t),\ldots,x_{P+s}(t)).\) We choose optimal values of~\(a_1\) and~\(a_2\) later for minimizing the redundancy. The encoding and decoding described in this subsection hold for all values of~\(a_1\) and~\(a_2.\)

We see below that the code~\({\cal C}_{far}\) is capable of correcting all~\(3P-\)far deletable patterns. The main idea is that if~\(\mathbf{x}\) is corrupted by a~\(3P-\)far deletable error pattern, then at most one deletable error occurs in each word~\(\mathbf{x}(j)\)
in~(\ref{code_def2}). This allows us to correct the errors in a sequential manner. Let~\(\mathbf{y} = F_{\mathbf{g}}(\mathbf{x})\) where~\(\mathbf{g} = (\mathbf{e},\mathbf{v})\) be the received word and  suppose that~\(\mathbf{x}(j), j \leq t-2\) is the first block of~\(\mathbf{x}\) to be corrupted by an error in~\(\mathbf{g};\) i.e.,~\(\mathbf{e} = (e_1,\ldots,e_n)\) is such that if~\(u = \min\{i : e_i = 1\},\) then~\((j-1)P +1 \leq u \leq jP.\) We consider the cases~\(j = t-1\) or~\(j=t\) at the end.

Divide the corrupted word~\(\mathbf{y} = (\mathbf{y}(1),\ldots,\mathbf{y}(t_1))\) into blocks where the~\(j^{th}\) block~\(\mathbf{y}(j) = (y_1(j),\ldots,y_P(j)), 1 \leq j \leq t_1-1\) has~\(P\) bits and~\(\mathbf{y}(t_1)\) has at least~\(P\) and at most~\(2P-1\) bits. Since the first error position~\((j-1)P +1 \leq u \leq jP\) falls in block~\(\mathbf{x}(j),\) the blocks~\(\mathbf{x}(i),1 \leq i \leq j-1\) are uncorrupted and~\(\mathbf{y}(i) = \mathbf{x}(i),1 \leq i \leq j-1.\) Consequently, the checksum
\begin{equation}\label{check_sum_i}
CS(l) := \sum_{i=1}^{P}i \cdot y_i(l) \equiv a_1 \mod(2P+1)
\end{equation}
for~\(1 \leq l \leq j-1\) and so the checksum difference~\(|CS(l) - a_1| \mod(2P+1) \equiv 0\) for all blocks~\(1 \leq l \leq j-1.\)

If the error corrupting the block~\(\mathbf{x}(j)\) is an erasure, then~\(y_u(j) = \varepsilon\) (the erasure symbol) and~\(y_k(j) = x_k(j)\) for all~\(1 \leq k \neq u \leq P.\) Therefore the modified checksum
\begin{equation}\label{er_chk_sum}
CS_{er}(j) := \sum_{i=1, i \neq u}^{P} i \cdot y_{i}(j) = \sum_{i=1, i \neq u}^{P} i \cdot x_i(j).
\end{equation}
If~\(x_{u}(j) =0,\) then~\(CS_{er}(j) - a_1 \equiv 0 \mod (2P+1)\) and if~\(x_{u}(j) =1,\) then\\\(CS_{er}(j) - a_1 \equiv -u \mod(2P+1) \neq 0\) and so the modified checksum difference~\(|CS_{er}(j) - a_1| \mod(2P+1) \equiv u \ind(x_u(j) = 1)\) is nonzero if and only if the erased bit is a one. This allows us to correct the erasure. %erased bit~\(x_{u}(j).\)

If the error corrupting the block~\(\mathbf{x}(j)\) is a substitution or deletion, then it is not directly detectable as in the case of erasure above. We therefore use the checksums of both~\(\mathbf{y}(j)\) and~\(\mathbf{y}(j+1)\) to indirectly deduce the nature of the error. Indeed, if the error is a substitution then~\(y_u(j) = 1-x_{u}(j)\) and~\(y_k(j) = x_k(j)\) for~\(1 \leq k \neq u \leq P\) and so the checksum of the block~\(\mathbf{y}(j)\) is
\begin{equation}\label{chk_flip}
CS(j) := \sum_{i=1}^{P} i \cdot y_{i}(j) = \sum_{i=1, i\neq u} i \cdot x_{i}(j) + u\cdot(1-x_{u}(j)).
\end{equation}
Thus~\(CS(j) - a_1 \equiv u(1-2x_{u}(j))\) and so the checksum difference for the block~\(\mathbf{y}(j),\) \[|CS(j)-a_1| \equiv u \mod(2P+1)\]  is nonzero. Moreover, since the errors are~\(3P-\)far apart, no errors have corrupted the block~\(\mathbf{x}(j+1)\) of the original word (which also contains~\(P\) bits since~\(j \leq t-2\)). Therefore the block~\(\mathbf{y}(j+1) = \mathbf{x}(j+1)\) and the checksum difference\\\(|CS(j+1)-a_1| \mod(2P+1)\) for the block~\(\mathbf{y}(j+1)\) is zero.

If on the other hand the error corrupting the block~\(\mathbf{x}(j)\) is a deletion, we may or may not get a nonzero checksum difference in block~\(\mathbf{y}(j).\) But we are guaranteed a nonzero checksum difference in block~\(\mathbf{y}(j+1).\) This is because in the block~\(\mathbf{y}(j+1),\) the corresponding bits of~\(\mathbf{x}(j+1)\) are shifted one position to the left; i.e.,~\(y_i(j) = x_{i+1}(j)\) for~\(1 \leq i \leq P-1.\) Therefore the checksum
\begin{eqnarray}
CS(j+1) &=& \sum_{i=1}^{P-1}i\cdot x_{i+1}(j+1) + P\cdot y_P(j+1) \nonumber\\
&=& \sum_{i=1}^{P-1} (i+1) \cdot x_{i+1}(j+1) + P\cdot y_P(j+1) - \sum_{i=2}^{P} x_i(j+1) \nonumber\\
&=& \sum_{i=1}^{P}i \cdot x_i(j+1) + P\cdot y_P(j+1) - \sum_{i=1}^{P} x_i(j+1) \nonumber
\end{eqnarray}
and the checksum difference~\[CS(j+1) - a_1 \equiv P\cdot y_P(j+1) - \sum_{i=1}^{P} x_i(j+1) \mod(2P+1).\] If~\(y_P(j+1) = 1,\) then~\(P \geq P - \sum_{i=1}^{P} x_i(j+1) \geq 1,\) since not all bits in a block can be one. If~\(y_P(j+1) = 0,\)
then~\(-P \leq - \sum_{i=1}^{P} x_i(j+1) \leq -1,\) since not all bits in a block can be zero. In other words, the checksum difference for the block~\(\mathbf{y}(j+1),\)~\(|CS(j+1)-a_1| \mod(2P+1) \neq 0.\)

%After detecting and determining whether the error is a deletion or flip as described above, we now perform the appropriate correction in~\(\mathbf{y}(j)\) to get back~\(\mathbf{x}(j).\) If the error is a deletion, then perform single deletion correction with the first~\(P-1\) bits~\(\{y_i(j)\}_{1 \leq i \leq P-1}\) to get a new~\(P\) bit word~\(\mathbf{\hat{x}}(j)\) which equals~\(\mathbf{x}(j).\) If the error is a flipping, then we know from discussion following~(\ref{chk_flip}) that~\(|CS(j)-a_1| \equiv k \mod {P+1}.\) This obtains the position of the substituted bit and allows us correct the error.

%If a deletion had occurred, then we perform single deletion correction on the first~\(P-1\) bits of~\(\mathbf{y}(j)\) to get a new word~\(\mathbf{\hat{w}}\)
%which necessarily equals~\(\mathbf{x}(j).\) This corrects the deletion in~\(\mathbf{x}(j).\)

%Letting~\(\mathbf{y}^{(1)}\)
%denote the new word, we then perform the same procedure as above on~\(\mathbf{y}^{(1)}.\)

%The word~\(\mathbf{y}^{(1)}\) obtained after error correction as described above contains one less error than~\(\mathbf{y}.\) Again if the first error in~\(\mathbf{y}^{(1)}\) occurs in a Repeating the above procedure iteratively, we correct a single error in every iteration.

Summarizing, suppose we get the first nonzero checksum difference (mismatch) at some block~\(\mathbf{y}(j)\) containing~\(P\) bits. The mismatch may be because of an error either in block~\(\mathbf{y}(j-1)\) or block~\(\mathbf{y}(j).\) If indeed the mismatch was because of an error in block~\(\mathbf{y}(j-1),\) then the error must be a deletion since a substitution would have already caused a mismatch for block~\(\mathbf{y}(j-1)\) (see discussion following~(\ref{chk_flip})). We therefore remove the last bit of the block~\(j-1\) and perform single deletion correction to get a new word~\(\mathbf{\hat{x}}(j-1).\) If~\(\mathbf{\hat{x}}(j-1) \neq \mathbf{y}(j-1),\) we replace the first~\(P-1\) bits of the block~\(\mathbf{y}(j-1)\) with the block~\(\mathbf{\hat{x}}(j-1).\) %and finish the error correct

If~\(\mathbf{\hat{x}}(j-1) = \mathbf{y}(j-1),\) then we can assume that the error has occurred in block~\(\mathbf{y}(j)\) and use block~\(\mathbf{y}(j+1)\) to determine the nature of error: If~\(\mathbf{y}(j+1)\) is not the final block in~\(\mathbf{y},\) then we proceed as above and deduce that the error in~\(\mathbf{y}(j)\) is either a substitution or deletion depending on whether the checksum difference of the block~\(\mathbf{y}(j+1)\) is zero or not, respectively.

If on the other hand~\(\mathbf{y}(j+1)\) is the final block in~\(\mathbf{y},\) then it is necessarily true that~\(\mathbf{y}\) is obtained after corrupting~\(\mathbf{x}\) by a single error occurring in one of the last two blocks~\(\mathbf{x}(t-1)\) or~\(\mathbf{x}(t);\) i.e., the first error position~\(u\) defined in the paragraph prior to~(\ref{check_sum_i}) falls in one of the last two blocks~\(\mathbf{x}(t-1)\) or~\(\mathbf{x}(t).\) Since errors are~\(3P-\)far apart, the error at position~\(u\) is the only error corrupting~\(\mathbf{x}\) and so the corrupted word~\(\mathbf{y}\) has either~\(n\) or~\(n-1\) bits.

Split~\(\mathbf{y} = (\mathbf{y}(1),\ldots,\mathbf{y}(t-1),\mathbf{y}(t))\) where the first~\(t-1\) blocks contain~\(P\) bits each and the final block~\(\mathbf{y}(t)\) has either~\(P+s\) or~\(P+s-1\) bits. One of the blocks in~\(\{\mathbf{y}(t-1),\mathbf{y}(t)\}\) has been corrupted by an error
and we perform correction as follows. If the error is an erasure, then we correct the erased bit simply by computing the checksum difference as described before (see discussion prior to~(\ref{er_chk_sum})). If there is no erasure but the received word~\(\mathbf{y}\) has~\(n\) bits, then the error is a substitution and we obtain the location of the substitution by computing the checksum difference for each of the blocks~\(\mathbf{y}(t-1)\) and~\(\mathbf{y}(t),\) as described in the paragraph containing~(\ref{chk_flip}).

Finally, if~\(\mathbf{y}\) has~\(n-1\) bits, then the error is a deletion and we determine the block where the deletion has occurred as follows. We perform single deletion correction on the first~\(P~-~1\) bits~\(\{\mathbf{y}_i(t-1)\}_{1 \leq i \leq P-1}\) of~\(\mathbf{y}(t-1)\) and get a new~\(P\) bit word~\(\mathbf{\hat{w}}.\) If~\(\mathbf{\hat{w}} \neq \mathbf{y}(t-1),\) then the deletion necessarily has occurred in the block~\(\mathbf{x}(t-1)\) of the original word~\(\mathbf{x}.\) We therefore finish the iteration by inserting~\(\mathbf{\hat{w}}\) into the word~\(\mathbf{y}\) giving the output~\(\mathbf{\hat{x}} = \left(\mathbf{y}(1),\ldots,\mathbf{y}(t-2),\mathbf{\hat{w}},\right.\)\(\left.y_P(t-1),\mathbf{y}(t)\right).\)

If on the other hand~\(\mathbf{\hat{w}} = \mathbf{y}(t-1),\) then we assume that the deletion has occurred
in the final block~\(\mathbf{x}(t)\) of the original word~\(\mathbf{x}.\)  We then perform single deletion correction on the~\(P+s-1\) bits of~\(\mathbf{y}(t)\) to get a new word~\(\mathbf{\hat{z}}\) containing~\(P+s\) bits and output~\(\mathbf{\hat{x}} = (\mathbf{y}(1),\ldots,\mathbf{y}(t-1),\mathbf{\hat{z}}).\)

%So far, we have discussed the case where there is a checksum mismatch in some block~\(\mathbf{y}(j)\) containing~\(P\) bits. If there is no such
%mismatch, then the received word~\(\mathbf{y}\) has either~\(n\) or~\(n-1\) bits and splitting~\(\mathbf{y}\) as in~(\ref{y_splt}), we perform the necessary error correction on the final block~\(\mathbf{y}(t)\) accordingly.
The procedure described above corrects a single error in the corrupted word~\(\mathbf{y}\) and after the iteration, the first error occurring at position~\(u\) in the error pattern~\(\mathbf{g}\) (see discussion prior to~(\ref{check_sum_i})) is corrected. The received word after the first iteration~\(\mathbf{y}^{(1)} = F_{\mathbf{g}(1)}(\mathbf{x})\) is a lesser corrupted version of~\(\mathbf{x}\) than~\(\mathbf{y}\) and~\(\mathbf{g}(1) = (\mathbf{e}(1),\mathbf{v})\) is an error pattern whose first error occurs strictly after the first error in~\(\mathbf{g};\) i.e., if~\(\mathbf{e}(1) = (e_1(1),\ldots,e_n(1))\) and~\(u_1 = \min\{i:e_i(1) = 1\},\) then~\(u_1 > u.\)

We now repeat the above procedure to correct the error at position~\(u_1\) and continuing iteratively, we sequentially correct all the remaining errors. The procedure stops if there is no checksum mismatch and we output the final word as our estimate of~\(\mathbf{x}.\) We summarize the above method as an Algorithm before the Appendices, where the function~\(BlockSplit(.,P)\) splits the input into blocks of length~\(P\) with the last block containing between~\(P\) and~\(2P-1\) bits.

From the description, we also see that the above algorithm is capable of correcting any~\(P-\)far error pattern in real-time with a delay of at most~\(4P.\) This is because knowing the corruption of the first~\(z+4P\) bits of the codeword~\(\mathbf{x}\) for any integer~\(1 \leq z \leq n-4P,\) we can determine the value of the bits~\(x_i,1 \leq i \leq z.\)

%CHNG BELOW +eT...

%For~\(P \geq 2\) let
%\begin{equation}\label{del_def2}
%\delta(P) := \frac{P+1}{2^{P-1}} \text{ and }\gamma(P) := \log\left(\frac{1+1/P}{1-\delta(P)}\right),
%\end{equation}
%so that~\(\delta(P)\) and~\(\gamma(P)\) converge to zero as~\(P \longrightarrow \infty.\)

\subsection*{Redundancy of the code~\({\cal C}_{far}\)}
To compute an upper bound on the redundancy of the code~\({\cal C}_{far},\) we choose~\(a_1\) and~\(a_2\) above such that
\begin{equation}\label{a1_a2_choose}
\#VT_{a_1}(P) \geq \frac{2^{P}}{2P+1} \text{ and } \#VT_{a_2}(s) \geq \frac{2^{P+s}}{2(P+s)+1}.
\end{equation}
Since we append together~\(t-1\) words each chosen from~\(VT_{a_1}(P) \setminus \{\mathbf{0},\mathbf{1}\}\) and one word from~\(VT_{a_2}(P+s),\) we have that
\begin{equation}
\#{\cal C}_{far} \geq \left(\frac{2^{P}}{2P+1}-2\right)^{t-1} \frac{2^{P+s}}{(2P+2s+1)} = 2^{n} \left(\frac{1-\delta(P)}{2P+1}\right)^{t-1} \frac{1}{2P+2s+1}, \label{cfar_1}
\end{equation}
where~\(\delta(P) = \frac{2P+1}{2^{P-1}}\) is as in~(\ref{del_def_main}) and relation~(\ref{cfar_1}) is true  since~\(Pt + s = n.\)
Using~\(s+1 \leq P,\) we further get
\[\#{\cal C}_{far} \geq 2^{n} \left(\frac{1-\delta(P)}{2P+1}\right)^{t-1} \frac{1}{4P} \geq 2^{n} \left(\frac{1-\delta(P)}{2P+1}\right)^{\frac{n}{P}-1} \frac{1}{4P},\]
since~\(t \leq \frac{n}{P}\) and~\(\frac{1-\delta(P)}{2P+1} < 1.\) From~(\ref{red_def}) we then get~(\ref{red_upp44}).

To obtain the redundancy lower bounds in Theorem~\ref{thm_11}, we use the fact that any code capable of correcting~\(t\) deletable errors must also be capable of corrected~\(t\) deletions. We also remark that redundancy lower bounds for segmented deletion channels have been studied in~\cite{rmj} and so for completeness, we give our proofs in the Appendix~\(1.\)

\setcounter{equation}{0}
\renewcommand\theequation{\arabic{section}.\arabic{equation}}
\section{Proof of Theorem~\ref{thm_12}}\label{pf_12}
Let
\begin{equation}\label{omega_cond}
\omega_n := \frac{4n}{t_n^2 \cdot d_n} \text{ and }P_n := \frac{n}{t_n^2 \omega_n} = \frac{d_n}{4}
\end{equation}
so that~\(6 \leq \omega_n \leq \frac{n}{t_n^3} \) and assume that~\(P_n\) is an integer greater than or equal to~\(6.\) Let~\({\cal C}_{frac} := {\cal C}_{far}(n,P_n)\) be the code defined in Theorem~\ref{thm_11} so that~\({\cal C}_{frac}\) is capable of real-time correction of all~\(3P_n-\)far deletable error patterns with a delay of at most~\(4P_n\) (see the final paragraph in the code description subsection of Section~\ref{p_far_codes}). Using~(\ref{red_upp44}),~\(P_n \geq 6\) and~\(\delta(P_n) \leq \frac{1}{2}\) we also get that the redundancy of~\({\cal C}_{frac}\)  is
\begin{equation}
R({\cal C}_{frac}) \leq \left(\frac{n}{P_n}-1\right) \log\left(\frac{2P_n+1}{1-\delta(P_n)}\right) + \log{P_n} + 2 \leq n \cdot \frac{\log{8P_n}}{P_n} \label{rqn1}
\end{equation}
where the second inequality in~(\ref{rqn1}) above follows from~\(P_n \geq 6.\) Setting~\(P_n = \frac{d_n}{4},\) we get the redundancy estimate~(\ref{red_upp}) in Theorem~\ref{thm_12}.

We show below that if~\({\cal F}_n\) is the set of all~\(3P_n-\)far error patterns, then
\begin{equation}\label{frac_pattern}
\frac{\#{\cal F}_n}{\#{\cal E}_n(t_n)} \geq 1-\frac{42}{\omega_n}.
\end{equation}
for all~\(n \geq 10.\) Since~\({\cal C}_{frac}\) is capable of correcting all~\(3P_n-\)far deletable error patterns, this proves the estimate on the fraction of deletable errors corrected in Theorem~\ref{thm_12}. In what follows we prove~(\ref{frac_pattern}). First, the total number of error patterns in~\({\cal E}_n(t_n)\) is~\(\#{\cal E}_n(t_n)  = \sum_{k=0}^{t_n} {n \choose k} 3^{k}.\)
Suppose~\(\mathbf{g} = (\mathbf{e},\mathbf{v}),\mathbf{e} = (e_1,\ldots,e_n)\) is not a~\(3P_n-\)far deletable error pattern. Let~\(n = 3LP_n + s, 0 \leq s \leq 3P_n-1\) and split~\(\mathbf{e} = (\mathbf{e}(1),\ldots,\mathbf{e}(L))\) into~\(L\) blocks, where the first~\(L-1\) blocks each have length~\(3P_n\) and the final block has length~\(3P_n+s.\) Since~\(\mathbf{g}\) is not a~\(3P_n-\)far error pattern, there are only two possibilities:\\
\((p1)\) One of the~\(L\) blocks in~\(\mathbf{e}\) contains at least two nonzero entries.\\
\((p2)\) Two consecutive blocks in~\(\mathbf{e}\) contain exactly one nonzero entry each.\\
We bound the number of patterns for each possibility.

To bound the number of patterns~\(\#{\cal F}(p1)\) in possibility~\((p1),\) we argue as follows: If~\({\cal S}(l), 1 \leq l \leq L\) is the set of error patterns~\(\mathbf{g} =(\mathbf{e},\mathbf{v})\) where the~\(l^{th}\) block~\(\mathbf{e}(l)\) of~\(\mathbf{e}\) has two or more nonzero entries, then
\begin{equation}\label{fp1_12}
\#{\cal F}(p1) \leq \sum_{l=1}^{L}\#{\cal S}(l) = (L-1) \#{\cal S}(1) + \#{\cal S}(L).
\end{equation}
In the Appendix~\(2\) below we use Binomial coefficient estimates to show that
\begin{equation}\label{sl_est}
\#{\cal S}(l) \leq \frac{84 P_n^2 t_n^2 + 8t_n^4}{n^2} (\#{\cal E}_n(t_n))
\end{equation}
for all~\(1 \leq l \leq L\) and since~\(L \leq \frac{n}{3P_n},\)
we get from~(\ref{fp1_12}) that
\begin{equation}
\#{\cal F}(p1) \leq \frac{n}{3P_n}\left(\frac{84P_n^2t_n^2 + 8t_n^4}{n^2}\right) (\#{\cal E}_n(t_n)) = \left(\frac{28P_nt_n^2}{n} + \frac{8t_n^4}{3nP_n}\right) (\#{\cal E}_n(t_n)). \nonumber
\end{equation}
Since~\(P_n = \frac{n}{t_n^2 \omega_n}\) we get that~\(\frac{28P_nt_n^2}{n} + \frac{4t_n^4}{3nP_n}  = \frac{28}{\omega_n} + \frac{4}{3\omega_n} \left(\frac{\omega_n t_n^3}{n}\right)^2 \leq \frac{28 + \frac{4}{3}}{\omega_n}\) by~(\ref{omega_cond}) and so
\begin{equation}\label{fp1_122}
\#{\cal F}(p1) \leq  \frac{30}{\omega_n} \cdot \left(\#{\cal E}_n(t_n)\right).
\end{equation}

To bound the number of patterns~\(\#{\cal F}(p2)\) in possibility~\((p2)\) described prior to~(\ref{fp1_12}), we argue as follows: If~\({\cal T}(l)\) is the set of error patterns~\(\mathbf{g} =(\mathbf{e},\mathbf{v})\) where the blocks~\(\mathbf{e}(l)\) and~\(\mathbf{e}(l+1),1 \leq l \leq L-1\) each have exactly one nonzero entry each, then
\begin{equation}\label{fp2}
\#{\cal F}(p2) \leq \sum_{l=1}^{L-1}\#{\cal T}(l) = (L-2) \#{\cal T}(1) + \#{\cal T}(L-1).
\end{equation}

We estimate~\(\#{\cal T}(1)\) as follows. Suppose~\(\mathbf{g} = (\mathbf{e},\mathbf{v}) \in {\cal T}(1)\) and suppose that~\(\mathbf{e} = (e_1,\ldots,e_n)\) has~\(k\) nonzero entries. We split~\(\mathbf{e} = (\mathbf{e}(1),\ldots,\mathbf{e}(L))\) into~\(L\) blocks as before. Among the~\(3P_n\) bits in the first block~\(\mathbf{e}(1),\) we choose one bit, among the~\(3P_n\) bits of~\(\mathbf{e}(2)\) we choose one bit and choose~\(k-2\) bits from the remaining~\(n-6P_n\) bits of~\(\mathbf{e}(j), j \neq 1,2.\) This can be done in~\({3P_n \choose 1}{3P_n \choose 1}{n-6P_n \choose k-2}\) ways and so
\begin{equation}
\#{\cal T}(1) = \sum_{k=2}^{t_n}  9P_n^2{n-6P_n \choose k-2} 3^{k} \leq 9P_n^2 \sum_{k=2}^{t_n} {n \choose k-2} 3^{k} . \label{t1_est1}
\end{equation}
For~\(k \leq t_n\) the ratio
\begin{equation}\label{nk_est1}
\frac{{n \choose k-2}}{{n \choose k}} = \frac{k(k-1)}{(n-k+2)(n-k+1)} \leq \frac{t_n^2}{(n-t_n)^2}=\left(\frac{t_n}{n}\right)^2 \left(1-\frac{t_n}{n}\right)^{-2}
\end{equation}
and using~\(\frac{t_n}{n} \leq \frac{1}{\sqrt{n}}\) (see statement of Theorem), we also have that~\(\left(1-\frac{t_n}{n}\right)^{-2} \leq \frac{n}{(\sqrt{n}-1)^2} \leq 4\) for all~\(n \geq 4.\)
From~(\ref{t1_est1}), we therefore get that
\begin{equation}
\#{\cal T}(1) \leq \frac{36P_n^2 t_n^2}{n^2}\sum_{k=2}^{t_n} {n \choose k} 3^{k} \leq \frac{36P_n^2 t_n^2}{n^2} (\#{\cal E}_n(t_n)). \label{t1_est12}
\end{equation}

Performing an analogous analysis as above for the last two blocks with length~\(3P_n\) and~\(3P_n+s,\) we get
\begin{equation}
\#{\cal T}(L-1) = \sum_{k=2}^{t_n}  3P_n(3P_n+s) {n-6P_n-s \choose k-2} 3^{k} \leq 18P_n^2 \sum_{k=2}^{t_n} {n \choose k-2} 3^{k}, \label{t2_est}
\end{equation}
since~\(s \leq 3P_n.\) Again using~(\ref{nk_est1}), we get
\begin{equation}
\#{\cal T}(L-1) \leq \frac{36P_n^2 t_n^2}{n^2}\sum_{k=2}^{t_n} {n \choose k} 3^{k} \leq \frac{36P_n^2 t_n^2}{n^2} (\#{\cal E}_n(t_n)). \label{t1_est121}
\end{equation}
Using~(\ref{t1_est12}) and~(\ref{t1_est121}) in~(\ref{fp2}) and we therefore get that
\begin{equation}\label{fp2_est}
\#{\cal F}(p2) \leq L \cdot \frac{36P_n^2 t_n^2}{n^2} (\#{\cal E}_n(t_n))  \leq \frac{12P_n t_n^2}{n} (\#{\cal E}_n(t_n)) =\frac{12}{\omega_n}  (\#{\cal E}_n(t_n)),
\end{equation}
where the second inequality in~(\ref{fp2_est}) is true since~\(L \leq \frac{n}{3P_n}\)  and the final estimate in~(\ref{fp2_est}) is true since~\(P_n =\frac{n}{t_n^2 \omega_n}\) (see~(\ref{omega_cond})).

Recall from discussion prior to~(\ref{fp1_12}) that~\({\cal F}_n\) denotes the set of all~\(3P_n-\)far error patterns and using~(\ref{fp1_122}) and~(\ref{fp2_est}), we get that~\(\frac{\#{\cal F}_n}{\#{\cal E}_n(t_n)} \geq 1-\frac{42}{\omega_n},\)
proving~(\ref{frac_pattern}).\;\;\;\;\;\;\;\;\;\;\;\;\;\;\;\;\;\;\;\;\;\;\;\;\;\;\;\;\;\;\;\;\;\;\;\;\;\;\;\;
\;\;\;\;\;\;\;\;\;\;\;\;\;\;\;\;\;\;\;\;\;\;\;\;\;\;\;\;\;\;\;\;\;\;\;\;\;\;\;\;\;\;\;\;\;\;\;\;\;\;\;\;\;\;\;\;\;\;\;\;\;\;\;\;\;\;~\(\qed\)

%\geq 1 - \frac{\#{\cal F}(p1) + \#{\cal F}(p2)}{\#{\cal E}_n(t_n)} \geq

%CHK ABV +eTC.. CRFLLY+EC...ANYWAYS...+eTC.. ALSO..HEH
\subsection*{Conclusion}
In this paper, we have studied redundancy of codes capable of correcting deletable errors in real time with low probability of error. We have also provided decoding algorithms that recover the errors in real time. It would be interesting to extend our results to other channel models, like for example, magnetic channels with possible additional constraints.

\subsubsection*{Acknowledgements}
I thank Professors Alberto Gandolfi, Federico Camia, C. R. Subramanian and the referees for crucial comments that led to an improvement of the paper. I also thank Professors Federico Camia, C. R. Subramanian and IMSc for my fellowships.

\bibliographystyle{plain}

\begin{thebibliography}{10}
\bibitem{abro} Abroshan, M., Venkataramanan, R. and Guill\'en i F\`abregas, A.,
\newblock{Coding for Segmented Edit Channels},
\newblock{\em IEEE Transactions on Information Theory}, \textbf{64}, 3086--3098 (2018).

\bibitem{al} Alon, N. and Spencer, J.,
\newblock{\em The Probabilistic Method},
\newblock{Wiley Interscience}.

%\bibitem{guru} Ganesan, G.,
%\newblock{Construction and Redundancy of Codes for Correcting Deletable Errors},
%\newblock{\em Arxiv Link: https://arxiv.org/pdf/1805.00776.pdf} (2018).

\bibitem{helb} {Helberg, A. S. J.  and Ferreira, H. C.},
\newblock {On Multiple Insertion/Deletable Error Correcting Codes},
\newblock {\em IEEE Transactions on Information Theory}, \textbf{48}, 305--308 (2002).

\bibitem{krishnan} Krishnan, M. N., Shukla, D. and Vijay Kumar, P.,
\newblock{Rate-Optimal Streaming Codes for Channels with Burst and Random Erasures},
\newblock{\em IEEE Transactions on Information Theory}, \textbf{66}, 4869--4891  (2020).

\bibitem{lev} Levenshtein, V. I.,
\newblock{Binary Codes Capable of Correcting Deletable Errors, Insertions and Reversals},
\newblock {\em Doklady Akademii Nauk SSSR}, \textbf{163}, 845--848 (1965).

\bibitem{liu}  Liu, Z. and Mitzenmacher, M.,
\newblock{Codes for Deletion and Insertion Channels with Segmented Errors},
\newblock{\em IEEE Transactions on Information Theory}, \textbf{56}, 224--232 (2010).



\bibitem{sima} Sima, J. and Bruck, J.,
\newblock{Optimal~\(k-\)Deletion Correcting Codes},
\newblock{\em Internation Symposium on Information Theory}, 847--851 (2019).

\bibitem{tat} Tatwawadi, K. and Chandak, S.,
\newblock{Tutorial on Algebraic Deletion Correction Codes},
\newblock{\em Arxiv Link: https://arxiv.org/pdf/1906.07887.pdf} (2019).

\bibitem{rmj} {Venkataramanan, R., Swamy, V. N. and Ramchandran, K.,}
\newblock{Low Complexity Interactive Algorithms for Synchronization from Deletions, Insertions, and Substitutions},
\newblock{\em IEEE Transactions on Information Theory}, \textbf{61}, 5670--5689 (2015).
\end{thebibliography}

\begin{algorithm}[tbp]\footnotesize
\caption{Correcting a~\(3P-\)far deletable error pattern}\label{euclid}
%\DontPrintSemicolon
\SetNoFillComment

\Input{Received word $\mathbf{y}$}
\Output{Estimated word $\mathbf{\hat{x}}$}%
\SequentialDecode{}
{
\emph{Initialization}: $\mathbf{\hat{x}} \gets \mathbf{y}, j \gets 0, CS  \gets 0.$ \\
\emph{Preprocessing}: $(\mathbf{\hat{x}}(1),\ldots,\mathbf{\hat{x}}(t-1),\mathbf{\hat{x}}(t)) \gets BlockSplit(\mathbf{\hat{x}},P),$\\
\emph{where $P \leq Length(\mathbf{\hat{x}}(t)) < 2P$ }.\\
\tcc{Pass all blocks with checksum match}
\While {$CS = 0$ and $j \leq t$ }{
    $j \gets j+1, CS \gets -1$\;
    \If{erasure in~$j^{th}$ block}
    {$\mathbf{\hat{x}}(j) \gets CorrectErasure(\mathbf{\hat{x}}(j))$\;}
    \If{$j < t$ or $j=t$ and $Length(\hat{\mathbf{x}}(t)) = P+s$}
        {$CS \gets CheckSumDifference(\mathbf{\hat{x}}(j))$\;}

    %\left(\sum_{i=1}^{P_{curr}}i \cdot \hat{x}_i(j) - a_{curr} \right) \mod (P_{curr}+1)$\;
}

%\tcp*{Go to previous block}\\

\If{$CS = 0$ and $j=t$ }
{
\textbf{output} $(\mathbf{\hat{x}}(1),\ldots,\mathbf{\hat{x}}(t-1),\mathbf{\hat{x}}(t))$\;
}

\tcc{Checksum mismatch in~\(j^{th}\) block; Check if deletion in~\((j-1)^{th}\) block}
$l \gets \min(j-1,1),\mathbf{T} \gets CorrectDeletion(RemoveLastBit(\mathbf{\hat{x}}(l)))$\; %\tcp*{Use VT decoder}\\
\eIf{$\mathbf{T} \neq \mathbf{\hat{x}}(l) $}  %\tcp{Error in~\(j^{th}\) block}
{$\mathbf{T} \gets (\mathbf{T},LastBit(\mathbf{\hat{x}}(l)))$\;}
{
\tcc{Error in~\(j^{th}\) block; If final block then directly perform correction}
\If{$j = t$}
{
\eIf{$\mathbf{\hat{x}}(j)$ has $P+s$ bits}
{$\mathbf{T} \gets CorrectFlip(\mathbf{\hat{x}}(j))$\;}
{$\mathbf{T} \gets CorrectDeletion(\mathbf{\hat{x}}(j))$\;}
{\textbf{output} $(\mathbf{\hat{x}}(1),\ldots,\mathbf{\hat{x}}(t-1),\mathbf{T})$\;}
}

\tcc{In all other cases, determine if error is flipping or deletion using block~\(j+1\)}
 $l \gets j, ErrorType \gets 1$\;
 \If{$l <t-1 $ or $l = t-1$ and $Length(\mathbf{\hat{x}}(l+1)) = P+s$}
 {$ErrorType \gets CheckSumDifference(\mathbf{\hat{x}}(l+1))$\;}
 \eIf{$ErrorType = 0$}
 {$\mathbf{T} \gets CorrectFlip(\mathbf{\hat{x}}(l))$\;}
 {$\mathbf{T} \gets CorrectDeletion(RemoveLastBit(\mathbf{\hat{x}}(l)))$\;
 $\mathbf{T} \gets (\mathbf{T},LastBit(\mathbf{\hat{x}}(l)))$\;
 }
}
%\EndIf

\tcc{Update the estimate~\(\mathbf{\hat{x}}\) with the new block $\mathbf{T}$}
{
$\mathbf{\hat{x}} \gets (\mathbf{\hat{x}}(1),\ldots,\mathbf{\hat{x}}(l-1),\mathbf{T},\mathbf{\hat{x}}(l+1),\ldots,\mathbf{\hat{x}}(t))$\;
\textbf{goto} \emph{Preprocessing}\;
}

%\EndIf

}
\end{algorithm}

\setcounter{equation}{0}
\renewcommand\theequation{\arabic{section}.\arabic{equation}}
\section*{Appendix 1: Redundancy lower bounds}
In our proofs below, we use the following large deviation estimate. Let~\(\{X_j\}_{1 \leq j \leq m}\) be independent Bernoulli random variables with~\[\mathbb{P}(X_j = 1) = p_j = 1-\mathbb{P}(X_j = 0)\] and
fix~\(0 < \epsilon  \leq \frac{1}{2}.\) If~\(T_m = \sum_{j=1}^{m} X_j\) and~\(\mu_m = \mathbb{E}T_m,\) then
\begin{equation}\label{conc_est_f}
\mathbb{P}\left(\left|T_m - \mu_m\right| \geq \mu_m \epsilon \right) \leq 2\exp\left(-\frac{\epsilon^2}{4}\mu_m\right)
\end{equation}
for all \(m \geq 1.\) For a proof of~(\ref{conc_est_f}), we refer to Corollary A.1.14, pp. 312 of~\cite{al}.

\emph{Proof of~(\ref{min_red2}) in Theorem~\ref{thm_11}}: Let~\({\cal D}\) be any code capable of correcting all~\(3P-\)far deletable error patterns.
The code~\({\cal D}\) is capable of correcting all~\(3P-\)far deletable error patterns consisting only of deletions
and so we henceforth consider only deletions. Let~\({\cal F}\) be the set
of all~\(3P-\)far deletable error patters consisting only of deletions  and for~\(\mathbf{x} \in {\cal D},\) set~\({\cal N}(\mathbf{x}) = \cup_{\mathbf{g} \in {\cal F}} F_{\mathbf{g}}(\mathbf{x}).\) Since the deletable errors are at least~\(3P\) apart, there are at most~\(\frac{n}{3P}\) deletable errors in~\(\mathbf{x}\) and so if~\(T\) is the largest integer less than or equal to~\(\frac{n}{3P},\) then~\({\cal N}(\mathbf{x}) \subseteq \cup_{0 \leq r \leq T} \{0,1\}^{n-r}.\) Moreover, if~\(\mathbf{x}_1 \neq \mathbf{x}_2 \in {\cal D},\) then~\(F_{\mathbf{e}}(\mathbf{x}_1) \neq F_{\mathbf{e}}(\mathbf{x}_2)\) by definition. So the sets~\(\{{\cal N}(\mathbf{x})\}\) are all disjoint and therefore
\begin{equation}\label{sum_less322}
\sum_{\mathbf{x} \in {\cal D}} \#{\cal N}(\mathbf{x}) \leq \sum_{r=0}^{T} 2^{n-r} = 2^{n+1}\left(1-\frac{1}{2^{T+1}}\right) \leq 2^{n+1}.
\end{equation}

In what follows, we estimate the size of~\({\cal N}(\mathbf{x})\) for each~\(\mathbf{x}\) belonging to a certain ``nice" subset of~\({\cal D}.\)  For~\(\mathbf{x} \in \{0,1\}^{n}\)
and integer~\(r \geq 1,\) let
\begin{equation}\label{xr_def2}
\mathbf{x}_r := \mathbf{x}_r(\mathbf{x}) = \mathbf{x}((3P+5)(r-1)+1,(3P+5)(r-1)+5)
\end{equation}
be a sequence of five bits of~\(\mathbf{x}.\) Say that~\(\mathbf{x}_r\) is a good~\(5-\)block if the
first and last bits are one and the remaining three bits are zero.
Let~\(Q(\mathbf{x})\) be the number of good~\(5-\)blocks in~\(\{\mathbf{x}_r\}_{1 \leq r \leq W}\)
where~\(W\) is the largest integer~\(x\) satisfying~\(x\cdot(3P+5) \leq n\) so that
\begin{equation}\label{we_22}
\frac{n}{3P+6} \leq \frac{n}{3P+5} -1 \leq W \leq \frac{n}{3P+5},
\end{equation}
for all~\(n\) large.

Defining
\begin{equation}\label{v_def2}
{\cal Y}_n  := \left\{\mathbf{x} \in \{0,1\}^{n} : Q(\mathbf{x}) \geq \frac{n}{64(3P+6)}\right\},
\end{equation}
we show at the end of this proof that
\begin{equation}\label{v_est2}
\#{\cal Y}_n \geq 2^{n}\left(1-\exp\left(-\frac{n}{16(3P+6)}\right)\right),
\end{equation}
for all~\(n\) large.

If~\(\mathbf{x} \in {\cal Y}_n \cap {\cal D},\) then there are at least~\(\frac{n}{64(3P+6)}\) good~\(5-\)blocks in~\(\mathbf{x}.\) Let~\(m_P\) be the largest integer less than or equal to~\(\frac{n}{64(3P+6)}.\) For every one of the~\(m_P\) good~\(5-\)blocks, we can either choose to remove a zero bit or not. There are~\(2^{m_P}\) ways of performing such a procedure and the resulting
set of corrupted words are all distinct. Therefore~\(\#{\cal N}(\mathbf{x}) \geq 2^{m_P}\)
and using~(\ref{sum_less322}) we get
\begin{equation}\nonumber
2^{m_P}\#\left({\cal D} \bigcap  {\cal Y}_n \right) \leq \sum_{\mathbf{x} \in {\cal D} \bigcap {\cal Y}_n} \#{\cal N}(\mathbf{x}) \leq 2^{n+1}.
\end{equation}
From~(\ref{v_est2}) we therefore get
\begin{equation}\nonumber
\#{\cal D} \leq \frac{2^{n+1}}{2^{m_P}} + \#{\cal Y}^c_n \leq \frac{2^{n+1}}{2^{m_P}} + 2^{n}\exp\left(-\frac{n}{16(3P+6)}\right) \leq \frac{2^{n+2}}{2^{m_P}}
\end{equation}
for all~\(n\) large, since~\(m_P \leq \frac{n}{64(3P+6)} \leq \frac{n\log{e}}{16(3P+6)}.\) Therefore the redundancy of~\({\cal D}\) is at least~\(m_P -\log{4} \geq \frac{n}{64(3P+6)}-3\) for all~\(n\) large, since~\(m_P \geq \frac{n}{64(3P+6)}-1.\)~\(\qed\)

\emph{Proof of~(\ref{v_est2})}: Let~\(\mathbf{X} = (X_1,\ldots,X_n)\) be a uniformly randomly chosen word in~\(\{0,1\}^{n}\)
so that~\(X_i, 1 \leq i \leq n\) are independent and identically distributed (i.i.d.)
with~\(\mathbb{P}(X_1 = 0) = \frac{1}{2} = \mathbb{P}(X_1 = 1).\)

To lower bound~\(Q(\mathbf{X}),\) let~\(r \geq 1\) be an integer and define
\begin{equation}\label{i1_def22}
I_r := \{X_{(3P+5)(r-1)+1} = 1\} \bigcap_{i=2}^{4} \{X_{(3P+5)(r-1)+i} = 0\} \bigcap \{X_{(3P+5)(r-1)+5} = 1\}
\end{equation}
be the event that~\(\mathbf{X}_r\) defined in~(\ref{xr_def2}) is a good block.
The events~\(\{I_r\}\) are i.i.d with~\(\mathbf{P}(I_1) = \frac{1}{32}\)
and so~\(Q(\mathbf{x}) = \sum_{r=1}^{W}I_r\) where~\(W\) is as in~(\ref{we_22})
has mean~\(\frac{W}{32}.\) Therefore using the deviation estimate~(\ref{conc_est_f}) with~\(\epsilon = \frac{1}{2}\)
we get
\begin{equation}\nonumber
\mathbb{P}\left(Q\left(\mathbf{x}\right) \geq \frac{W}{64}\right) \geq 1-\exp\left(-\frac{W}{16}\right).
\end{equation}
Using~(\ref{we_22}) we have that~\(\frac{W}{64} \geq \frac{n}{64(3P+6)}\) for all~\(n\) large and so
\begin{equation}\nonumber
\mathbb{P}\left(Q\left(\mathbf{x}\right) \geq \frac{n}{64(3P+6)}\right) \geq 1-\exp\left(-\frac{n}{16(3P+6)}\right)
\end{equation}
for all~\(n\) large.~\(\qed\)

\emph{Proof of~(\ref{min_red2}) in Theorem~\ref{thm_11}}: Let~\({\cal D}\) be any code capable of correcting all~\(3P_n-\)far deletable error patterns.
The code~\({\cal D}\) is capable of correcting all~\(3P_n-\)far deletable error patterns consisting only of deletions
and so we henceforth consider only deletions.
Let~\({\cal F}\) be the set of all~\(3P_n-\)far deletable error patters consisting only of deletions and for~\(\mathbf{x} \in {\cal D},\) set~\({\cal N}(\mathbf{x}) = \cup_{\mathbf{e} \in {\cal F}} F_{\mathbf{e}}(\mathbf{x}).\) Since the deletions are at least~\(3P_n\) apart, there are at most~\(\frac{n}{3P_n}\) deletable errors in~\(\mathbf{x}\) and so if~\(L\) is the largest integer less than or equal to~\(\frac{n}{3P_n},\) then~\({\cal N}(\mathbf{x}) \subseteq \cup_{0 \leq r \leq L} \{0,1\}^{n-r}.\) Moreover, if~\(\mathbf{x}_1 \neq \mathbf{x}_2 \in {\cal D},\) then~\(F_{\mathbf{e}}(\mathbf{x}_1) \neq F_{\mathbf{e}}(\mathbf{x}_2)\) by definition. So the sets~\(\{{\cal N}(\mathbf{x})\}\) are all disjoint and therefore
\begin{equation}\label{sum_less3}
\sum_{\mathbf{x} \in {\cal D}} \#{\cal N}(\mathbf{x}) \leq \sum_{r=0}^{L} 2^{n-r} = 2^{n+1}\left(1-\frac{1}{2^{L+1}}\right) \leq 2^{n+1}.
\end{equation}

As before, we estimate the size of~\({\cal N}(\mathbf{x})\) for a nice subset of~\({\cal D}.\)
Let~\(W\) be the largest integer~\(x\) such that~\(x\cdot 3P_n \leq n\) so that
\begin{equation}\label{we_2}
\frac{n}{3P_n}-1 \leq W \leq \frac{n}{3P_n}.
\end{equation}
For integer~\(1 \leq r \leq W\) and~\(\mathbf{x} \in \{0,1\}^{n},\) let
\begin{equation}\label{xr_def}
\mathbf{x}_r := \mathbf{x}_r(\mathbf{x}) = \mathbf{x}(3P_n(r-1)+1,3rP_n)
\end{equation}
be a sequence of~\(3P_n\) bits of~\(\mathbf{x}.\) We have thus divided the first~\(3WP_n\) bits of~\(\mathbf{x}\) into~\(W\) disjoint blocks~\(\{\mathbf{x}_r\}_{1 \leq r \leq W},\) each containing~\(3P_n\) bits.

Let~\(U\) be the largest integer~\(y\) such that~\(5 \cdot y \leq 3P_n\)
so that~\(\frac{3P_n}{5}-1 \leq U \leq \frac{3P_n}{5}.\) Divide the first~\(5U\) bits of each block~\(\mathbf{x}_r\) into disjoint blocks~\(\{\mathbf{x}_{r,j}\}_{1 \leq j \leq U}\) of five bits each. Say that~\(\mathbf{x}_{r,j}\) is a good~\(5-\)block if the first and last bits are one and the remaining three bits are zero.
Let~\(T_{r}(\mathbf{x})\) be number of good~\(5-\)blocks in~\(\mathbf{x}_r\) and  let
\begin{equation}\label{v_def}
{\cal V}_n := \left\{\mathbf{x} \in \{0,1\}^{n} : T_r(\mathbf{x}) \geq \frac{3P_n}{64} \text{ for all } 1 \leq r \leq W\right\}.
\end{equation}
At the end of this proof, we prove that there are positive constants~\(\beta_i,i=1,2\) such that
for all integers~\(n \geq \beta_1,\)
\begin{equation}\label{v_est}
\#{\cal V}_n \geq 2^{n}(1-e^{-\beta_2 P_n}).
\end{equation}

Let~\(W\) be as in~(\ref{we_2}) and as before, split the first~\(3WP_n\) bits of~\(\mathbf{x}\) into~\(W\) blocks~\(\{\mathbf{x}_r\}_{1 \leq r \leq W}\) of length~\(3P_n\) each. If~\(\mathbf{x} \in {\cal V}_n \cap {\cal D},\) then there at least~\(\frac{3P_n}{64}\) good~\(5-\)blocks in each~\(\mathbf{x}_r.\)
Choosing one good~\(5-\)block from each of~\(\mathbf{x}_1,\mathbf{x}_3,\ldots\) and deleting a middle zero from each such good block, we get a set of distinct words. Since there are at least~\(\frac{W-1}{2}\) blocks in~\(\mathbf{x}_1,\mathbf{x}_3,\ldots,\) the number of corrupted words obtained as above is at least~\(\left(\frac{3P_n}{64}\right)^{(W-1)/2}.\) The deletable error patterns giving rise to these words are all~\(3P_n-\)far apart and so we get
\begin{equation}\label{nx_f}
\#{\cal N}(\mathbf{x}) \geq \left(\frac{3P_n}{64}\right)^{(W-1)/2} \geq \left(\frac{3P_n}{64}\right)^{\frac{n}{6P_n}-1},
\end{equation}
since~\(W \geq \frac{n}{3P_n}-1,\) (see~(\ref{we_2})).

Letting~\[\Delta = \left(\frac{n}{6P_n}-1\right)\log\left(\frac{3P_n}{64}\right)\] and using~(\ref{nx_f}) in~(\ref{sum_less3}) we get
\begin{equation}\nonumber
2^{\Delta}\#\left({\cal D} \bigcap  {\cal V}_n \right) \leq \sum_{\mathbf{x} \in {\cal C} \bigcap {\cal V}_n} \#{\cal N}(\mathbf{x}) \leq 2^{n+1}
\end{equation}
and so from~(\ref{v_est}) we get
\begin{equation}\nonumber
\#{\cal D} \leq \frac{2^{n+1}}{2^{\Delta}} + \#{\cal V}^c_n \leq \frac{2^{n+1}}{2^{\Delta}} + 2^{n}e^{-\beta_2 P_n} \leq \frac{2^{n+2}}{2^{\Delta}}
\end{equation}
for all~\(n\) large, since using the second relation in~(\ref{pn_cond}) that~\(\frac{P_n^2}{n\log{n}} \longrightarrow 0\) as~\(n \rightarrow \infty,\) we have
\[\Delta \leq \frac{n}{P_n}\log(3P_n) \leq \frac{n \log{n}}{P_n} \leq \beta_2 P_n\log{e}\] for all~\(n\) large. Therefore the redundancy of~\({\cal D}\) is at least~\(\Delta -\log{4}\) for all~\(n\) large. This proves the redundancy lower bounds with the exception of~(\ref{v_est2}) and~(\ref{v_est}), which we prove separately below.~\(\qed\)

\emph{Proof of~(\ref{v_est})}: Let~\(\mathbf{X} = (X_1,\ldots,X_n)\) be a uniformly randomly chosen word in~\(\{0,1\}^{n}\)
so that~\(X_i, 1 \leq i \leq n\) are independent and identically distributed (i.i.d.)
with~\(\mathbb{P}(X_1 = 0) = \frac{1}{2} = \mathbb{P}(X_1 = 1).\)

To lower bound~\(T_r(\mathbf{X})\) for~\(1 \leq r \leq W,\) let~\(I_{r,j}\) be the event that~\(\mathbf{X}_{r,j}\) is a good~\(5-\)block.
The events~\(\{I_{r,j}\}_{1 \leq j \leq U}\) are i.i.d with~\(\mathbf{P}(I_{r,j}) = \frac{1}{32}\)
and so~\(T_r(\mathbf{X}) = \sum_{j=1}^{U} I_{r,j}\) has mean~\(\frac{U}{32}.\) Therefore from the deviation estimate~(\ref{conc_est_f}),
we get that
\begin{equation}\label{q_dev1}
\mathbb{P}\left(T_r\left(\mathbf{X}\right) \geq \frac{U}{64}\right) \geq 1-e^{-C_1U}
\end{equation}
for all~\(n\) large and some constant~\(C_1 > 0.\) Since~\(U \geq \frac{3P_n}{5} -1 \geq \frac{3P_n}{8}\) for all large~\(n,\)
we get
\begin{equation}\label{q_dev2}
\mathbb{P}\left(T_r\left(\mathbf{X}\right) \geq \frac{3P_n}{2^{9}}\right) \geq 1-e^{-4C_2P_n}
\end{equation}
for some constant~\(C_2 >0\) and all large~\(n.\) The constant~\(C_2\) does not depend on~\(r\) and so
\begin{equation}\label{q_dev3}
\mathbb{P}\left(\bigcap_{1 \leq r \leq W} \left\{T_r\left(\mathbf{X}\right) \geq \frac{3P_n}{2^{9}}\right\}\right) \geq 1-W\cdot e^{-4C_2P_n}.
\end{equation}
Using~(\ref{we_2}) and the first relation in~(\ref{pn_cond}) that~\(\frac{P_n}{\log{n}} \longrightarrow \infty,\) we get that~\[W\cdot e^{-4C_2P_n} \leq \frac{n}{2P_n} \cdot e^{-4C_2 P_n} \leq e^{-2C_2 P_n}\] for all~\(n\) large.~\(\qed\)

\setcounter{equation}{0}
\renewcommand\theequation{\arabic{section}.\arabic{equation}}
\section*{Appendix 2: Proof of~(\ref{sl_est})}
We estimate~\(\#{\cal S}(1)\) as follows. Suppose~\(\mathbf{g} = (\mathbf{e},\mathbf{v}) \in {\cal S}(1)\) and suppose that~\(\mathbf{e} = (e_1,\ldots,e_n)\) has~\(k \leq t_n\) nonzero entries. We split~\(\mathbf{e} = (\mathbf{e}(1),\ldots,\mathbf{e}(L))\) into~\(L\) blocks as before. Among the~\(3P_n\) bits in the first block~\(\mathbf{e}(1),\) we choose~\(2 \leq r \leq k\) of the bits in~\({3P_n \choose r}\) ways and choose~\(k-r\) bits of the remaining~\(n-3P_n\) bits in~\({n-3P_n \choose k-r}\) ways. The binomial coefficient~\({3P_n \choose r}\)  is nonzero since~\(r \leq k \leq t_n\) and~\(P_n = \frac{n}{t_n^2 \omega_n} \geq t_n\) (see~(\ref{omega_cond})). Thus~\(\#{\cal S}(1) = \sum_{k=2}^{t_n} \sum_{r=2}^{k} {n-3P_n \choose k-r} {3P_n \choose r} 3^{k} \)
and using~\(\sum_{k=0}^{r} {x \choose k-r}{y \choose r} = {x+y \choose k}\) with~\(x = n-3P_n\) and~\(y = 3P_n,\) we get
\begin{equation}\label{fp11}
\#{\cal S}(1) = \sum_{k=2}^{t_n} \left(1-S(n,k)\Delta\right) {n \choose k} 3^{k} ,
\end{equation}
where~\(S(n,k) = \frac{{n-3P_n \choose k}}{{n \choose k}}\) and
\begin{equation}
\Delta = 1 + \frac{{3P_n \choose 1} \cdot {n-3P_n \choose k-1}}{{n-3P_n  \choose k}} = 1 + \frac{3P_nk}{n-3P_n-k+1} \geq 1 + \frac{3P_n k}{n}. \label{del1del2}
\end{equation}

We now evaluate~\(S(n,k)\) by letting~\(x = n-3P_n\) to get
\begin{equation}\label{snk_def2}
S(n,k) = \frac{{x \choose k}}{{n \choose k}}=  \frac{x(x-1)\ldots(x-k+1)}{n(n-1)\ldots(n-k+1)} = \left(\frac{x}{n}\right)^{k}\frac{L(x,k)}{L(n,k)},
\end{equation}
where~\(1 \geq L(x,k) := \prod_{i=0}^{k-1} \left(1-\frac{i}{x}\right) \geq \left(1- \frac{1}{x}\sum_{i=1}^{k-1}i\right)= 1-\frac{k(k-1)}{2x}.\)
The term~\(x = n-3P_n  = n -\frac{3n}{t_n^2 \omega_n} \geq n - \frac{3n}{\omega_n} \geq \frac{n}{2}\) since~\(\omega_n \geq 6\) (see statement of Theorem) and so
\[1-\frac{k(k-1)}{2x} = 1- \frac{k(k-1)}{2n} - \frac{3P_nk(k-1)}{2nx} \geq 1- \frac{k(k-1)}{2n} - \frac{3P_nk^2}{n^2}\]
and
\begin{equation}\label{lx_est1}
L(x,k) \geq 1- \frac{k(k-1)}{2n} - \frac{3P_nk^2}{n^2}.
\end{equation}

To find an upper bound for~\(L(n,k),\) we use the fact that~\(1-y <e^{-y}\) for\\\(0 < y< 1\) to get~\(\log{L(n,k)} \leq -\sum_{i=1}^{k-1} \frac{i}{n}  = -\frac{k(k-1)}{2n}.\) But~\(\frac{k(k-1)}{2n} \leq \frac{t_n^2}{2n} \leq \frac{1}{2}\) since~\(t_n \leq \sqrt{n}\) (see statement of Theorem) and so using~\(e^{-z} \leq 1-z+z^2\) for~\(z < \frac{1}{2},\) we get
\begin{equation}
L(n,k) \leq \exp\left(-\frac{k(k-1)}{2n}\right) \leq 1- \frac{k(k-1)}{2n} + \left(\frac{k(k-1)}{2n}\right)^2  \leq 1- \frac{k(k-1)}{2n} + \frac{k^4}{n^2}.\label{lnk_up}
\end{equation}
Using~(\ref{lnk_up}) and the lower bound~(\ref{lx_est1}) we have
\begin{equation} \label{l_rat}
\frac{L(x,k)}{L(n,k)} -1 \geq \frac{1-\frac{k(k-1)}{2n} - \frac{3P_nk^2}{n^2}}{1- \frac{k(k-1)}{2n} + \frac{k^4}{n^2}} - 1  = -\frac{3P_nk^2 + k^4}{n^2}\left(1- \frac{k(k-1)}{2n} + \frac{k^4}{n^2}\right)^{-1}
\end{equation}

Since~\(k \leq t_n \leq \sqrt{n}\) (see statement of Theorem) we have~\[1- \frac{k(k-1)}{2n} + \frac{k^4}{n^2} \geq 1-\frac{t_n(t_n-1)}{2n}\geq \frac{1}{2}\] and so we get from~(\ref{l_rat}) that~\(\frac{L(x,k)}{L(n,k)} -1 \geq -2\frac{3P_nk^2 + k^4}{n^2}.\)
Substituting this estimate into~(\ref{snk_def2}) and using~\(\left(1-\frac{3P_n}{n}\right)^{k} \geq 1-\frac{3P_nk}{n}\) we get
\begin{equation}
S(n,k) \geq \left(1-\frac{3P_nk}{n}\right)\left(1-2\frac{3P_nk^2 + k^4}{n^2}\right) \geq 1-\frac{3P_n k}{n} - 2\frac{3P_nk^2 + k^4}{n^2}. \label{phjj_111}
\end{equation}
Using~(\ref{phjj_111}) and~(\ref{del1del2}), we therefore get that~\(S(n,k)\cdot \Delta\) is bounded below by
\begin{equation}
1-\left(\frac{3P_n k}{n}\right)^2 -2\frac{3P_nk^2 + k^4}{n^2}\left(1 + \frac{3P_n k}{n}\right). \nonumber
\end{equation}
Since~\(k \leq t_n\) we have that~\(\frac{P_n k}{n} \leq \frac{P_n t_n}{n} = \frac{1}{t_n \omega_n} \leq 1,\) because~\(\omega_n\) and~\(t_n\) are both at least one. Thus~\(1+\frac{3P_n k}{n} \leq 4\)
and so~\[S(n,k)\Delta \geq 1-\frac{9P_n^2k^2}{n^2} - 8 \frac{3P_nk^2 + k^4}{n^2}.\] Substituting this into~(\ref{fp11}) and using~\(k \leq t_n\) we get~\[\#{\cal S}(1) \leq \theta \cdot \sum_{k=2}^{t_n}{n \choose k} 3^{k} \leq \theta \cdot \#{\cal E}_n(t_n),\] where~\(\theta := \frac{9P_n^2t_n^2}{n^2} + 8 \frac{3P_nt_n^2 + t_n^4}{n^2}.\)
The above estimate holds for all the sets~\({\cal S}(l), 1 \leq l \leq L-1\) (see discussion prior to~(\ref{fp1_12})). Performing the same analysis with~\(3P_n+s\) instead of~\(3P_n\) we get \[\#{\cal S}(L) \leq \left(\frac{(3P_n+s)^2t_n^2}{n^2} + 8\frac{(3P_n+s)t_n^2 + t_n^4}{n^2}\right) (\#{\cal E}_n(t_n))\] and using the fact that~\(s \leq 3P_n,\) we therefore get
\[\#{\cal S}(l) \leq \left(\frac{(6P_n)^2t_n^2}{n^2} + 8 \frac{6P_nt_n^2 + t_n^4}{n^2}\right) (\#{\cal E}_n(t_n)) \leq \frac{84 P_n^2 t_n^2 + 8t_n^4}{n^2} (\#{\cal E}_n(t_n)) \]
for all~\(1 \leq l \leq L\)

\end{document}